# Symmetry and nonlinearity of spin wave resonance excited by focused surface acoustic waves

## Authors

Piyush J. Shah[1], Derek A. Bas[1], Abbass Hamadeh[2], Michael Wolf[1], Andrew Franson[1], Michael Newburger[1], Philipp Pirro[2], Mathias Weiler[2], Michael R. Page[1*]

## Affiliations

[1] Materials and Manufacturing Directorate, Air Force Research Laboratory, Wright-Patterson Air Force Base, Ohio 45433, USA
[2] Fachbereich Physik and Landesforschungszentrum OPTIMAS, Technische Universität Kaiserslautern, 67663 Kaiserslautern, Germany

[*]Email: michael.page.16@us.af.mil

## Abstract

The use of a complex ferromagnetic system to manipulate GHz surface acoustic waves is a rich current topic under investigation, but the high-power nonlinear regime is under-explored. We introduce focused surface acoustic waves, which provide a way to access this regime with modest equipment. Symmetry of the magneto-acoustic interaction can be tuned by interdigitated transducer design which can introduce additional strain components. Here, we compare the impact of focused acoustic waves versus standard unidirectional acoustic waves in significantly enhancing the magnon-phonon coupling behavior. Analytical simulation results based on modified Landau-Lifshitz-Gilbert theory show good agreement with experimental findings. We also report nonlinear input power dependence of the transmission through the device. This experimental observation is supported by the micromagnetic simulation using mumax3 to model the nonlinear dependence. These results pave the way for extending the understanding and design of acoustic wave devices for exploration of acoustically driven spin wave resonance physics.

## Introduction

Over the last decade, the magneto-acoustic interaction has become an important phenomenon of study for enhanced magnonic interactions and novel strain mediated architectures. Unlike other methods of spin manipulation which often rely on large currents or external fields, magneto-acoustic devices [1]–[5] rely on phonons which are commonly generated through voltage or optical schemes, and then converted into magnons or spin currents through localized absorption. This transduction scheme offers a compact, efficient method of probing and controlling magnetic domains and excitations. Further understanding and engineering of this interaction is of fundamental importance and will have an impact in the fields of spintronics; magnetic sensing; non-volatile, high-capacity memory; novel materials design; and frequency agile communications devices [6], [7].

Among the scope of work in this area, particular interest has been paid to the interaction of surface acoustic waves (SAWs) with magnetic thin films [8]–[13]. Most of this work is based on the established technique of using metal-electrode interdigitated transducers (IDTs) to generate propagating SAWs in a piezoelectric material. The frequency of the SAWs is tunable by a host of factors including the acoustic velocity of the piezoelectric and pitch of the IDTs, which enables frequencies in the GHz regime, complementary to those for spin waves and magnetic resonance. Thus, when a magnetostrictive thin film is in contact with the piezoelectric, the SAWs are converted into a local field at the interface which drives magnetization dynamics. Considerable work has been focused on identification and integration of magnetic materials which exhibit large magnetostriction and low damping to enhance the acousto-magnetic conversion [14], as well as piezoelectric materials with improved strain and frequency characteristics. Yet even the most basic systems consisting of Ni thin films on $LiNbO_3$ substrates are rich in physics, and work in this field has revealed multiple exciting behaviors including nonreciprocity of phase and amplitude [15]–[17], and stimulation of ferromagnetic resonance or spin wave modes. Although previous publications have often used the term acoustically-driven ferromagnetic resonance (ADFMR) to refer to the scope of dynamic magnetic behavior, we will instead refer to the methods discussed here as acoustically-driven spin wave resonance (ADSWR) to avoid the implication of uniform magnetic precession which is not a necessary condition [18]–[20].

Despite the expanding scope of work that exists in this field, the effect of increased elastic strain on magnon-phonon coupling has largely remained an under-explored topic. In particular, conventional geometries are limited by the piezoelastic constant of the substrate or the amount of power that can be applied before breakdown of the device. However, an alternate mechanism for enhancing the strain exists though focusing of the surface acoustic waves. This can be accomplished by curving the IDT pairs to generate a focusing wavefront. These focused IDTs (FIDTs) can be tuned through changes in the curvature and arc length, and work in these devices has proven beneficial to a number of fields including acoustofluidics for manipulation of cells; particle sorting; particle concentration and droplet production [19]–[23], [24]; manipulation of electron-hole pairs in GaAs quantum wells; and even in magnetization switching in FeGa thin film [27], where the application of focused SAWs lowers the coercive field of the magnet.

In this study, we report on the enhanced magneto-acoustic interaction in ADSWR devices achieved through focusing of acoustic waves. We study the impact of FIDT design by comparing the performance of traditional straight-finger IDT SAW delay line device to FIDT devices. We observe enhanced acoustic absorption from the FIDT geometry at a particular angle that is dependent on the arc length, along with substantial nonlinearity beginning at modest threshold input powers in the milliwatt range. We also provide analytical simulation results based on modified Landau-Lifshitz-Gilbert theory, which show good agreement with experimental findings. In addition, we also report on the nonlinear input power dependence of the transmission through the device which is supported by the micromagnetic simulation using mumax3 to model the nonlinear dependence, which is an important finding that will have benefit to a variety of other results.

The device geometry for our study involves a split-finger focused IDT design for Rayleigh wave generation in single crystal $y$-cut $LiNbO_3$ substrate. Favorable SAW propagation is along the $z$-axis between two pairs of IDTs for the delay line filter geometry as shown in FIG. 1A. The split-finger design minimizes the destructive interference caused by reflection from the IDTs and thereby allows device operation at higher odd harmonics of the fundamental frequency. The fundamental frequency $f_1$ of our design is around 291 MHz, however, most of the reported measurements are taken at the 3$^{rd}$ harmonic $f_3$, ~873 MHz. Higher harmonics would be preferable as typical spin wave resonances tend to be above 1 GHz, but low-loss production of acoustic waves remains an engineering challenge due to impedance mismatch. In the current study we design devices with three focusing angles $\theta = 0°$, $45°$, $60°$, as shown in FIG. 1A, where $\theta = 0°$ refers to the standard straight IDT geometry. We chose the focal length $F_l = 800$ μm defined as the radius of the IDT arc, which is also used to set the distance of each IDT from the center of the ferromagnetic thin film as shown in FIG. 1B. Results for focal length $F_l = 400$ μm are included in the Supplemental Material, demonstrating a shorter magnetoelastic interaction length which scales the absorption accordingly, but otherwise qualitatively similar behavior. For systematic side-by-side comparison, the FIDT device and straight finger IDT device were both fabricated on the same die. The IDTs are patterned out of 70 nm Al electrodes while the magnetic film is 20 nm of Ni deposited by e-beam evaporation. Details of the fabrication process can be found in the Supplemental Material. Results reported in this study are from devices with $N_p = 10$ finger pairs with the minimum electrode separation $\lambda/8 = 1.5$ µm.

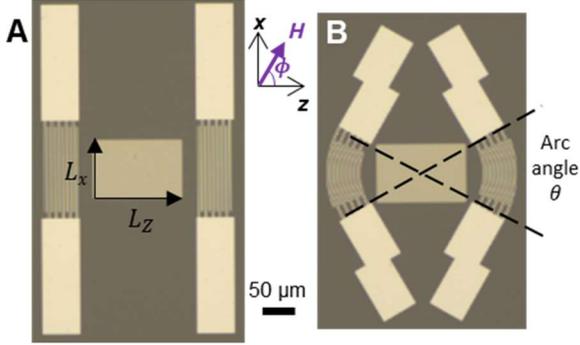

FIG. 1 Optical micrograph of example devices with **A** straight IDTs and **B** focused IDTs. The substrate is $y$-cut LiNbO$_3$, and the SAWs travel along the $z$-axis. The magnets used in this work have dimensions $L_x = 800$ µm, $L_z = 1225$ µm. Insets show the coordinate system with direction $\phi$ of applied field $H$, and 50-µm scale bar.

In FIG. 2 we show a microwave transmission polar plot for a typical ADSWR device. In these measurements, the SAW propagation direction and hence, k-vector, is kept fixed while the magnitude and direction of the applied magnetic field is varied using a vector electromagnet. The field is consistently swept from high magnitude (5 mT) to low magnitude (0 mT) to ensure the magnet begins in a saturated state. Additionally, we pulse the RF signal from the function generator and time-gate the measurement to isolate the signal from the SAWs, which arrives around 1 µs later than the spurious signal from local EM radiation. In these plots, red (blue) indicate the maximum (minimum) transmission respectively. The ADSWR-related absorption contrast then is defined as the difference between minimum and maximum signal for a given input RF power. Notably, the signals generally demonstrate a four lobe pattern of absorption which is the result of magneto-acoustic attenuation by the magnet. For this figure, RF input power was near 0 dBm and the observed contrast was constant with changes in power (see FIG. 3 for power dependence).

In FIG. 2A, for travelling wave straight finger IDT, the ADSWR contrast for the low RF input power case is about 3.4 dB (2.8 dB/mm), which agrees with previous studies at this frequency [28], [29] and shows the four-lobe symmetry with maximum absorption at about 20° from the SAW transmission axis (horizontal axis in the figures). In the simplest cases, this angle is about 45°, with any shifts generally attributed to magnetic anisotropy [5], [30]. As expected for the traditional IDT geometry, these experimental results match quite well with analytical model using Landau-Lifshitz-Gilbert based theory as introduced above, which we use to predict the symmetry of the output (impact of FIDT angle in the low-power limit). The micromagnetic simulations using mumax3 also match the experimental results, and we use these simulations to predict nonlinear power dependence. All parameters used in the analytical model and the micromagnetic simulations can be found in the Supplemental Material. Transmission plots in FIG. 2B and C show the influence of focusing SAWs on the center of the ferromagnetic film, with an IDT arc length θ of 45° and 60°, respectively. In FIG. 2B (θ = 45°) we observe a rotation of the lobes with an angular shift in the maximum absorption location to -5 and 80°. The simulation captures the shift accurately, except for the relative strength of the lobes (8.5 and 6.1 dB for the -5 and 80° lobes, respectively). Further rotation is observed in Fig. 2C ($\theta = 60°$), and the contrast changes further to 11.3 and 3.3 dB for lobes at -5 and 135°. The increase in the focusing angle affects the absorption symmetry in the device and increases the absorption contrast along two directions. Both the analytical model and the micromagnetic simulation reproduce the salient features of these experimental observations. This implies that – in the linear regime – the dominant effect of using FIDTs instead of straight IDTs is a change in the symmetry of the effective magnetoacoustic driving field acting on the Ni magnetization. FIDTs allow to efficiently excite spin-waves in Damon-Eshbach ($\phi = 90°$) or backward volume $\phi = 0°$ geometry on LiNbO$_3$ substrate, while Rayleigh-waves excited by straight IDTs on LiNbO$_3$ can only excite spin-waves efficiently at $\phi = 45°$. In this sense, FIDT are an alternative approach to using different piezoelectric substrates and thus SAW modes [9] to change the symmetry of the magnetoelastic interaction.

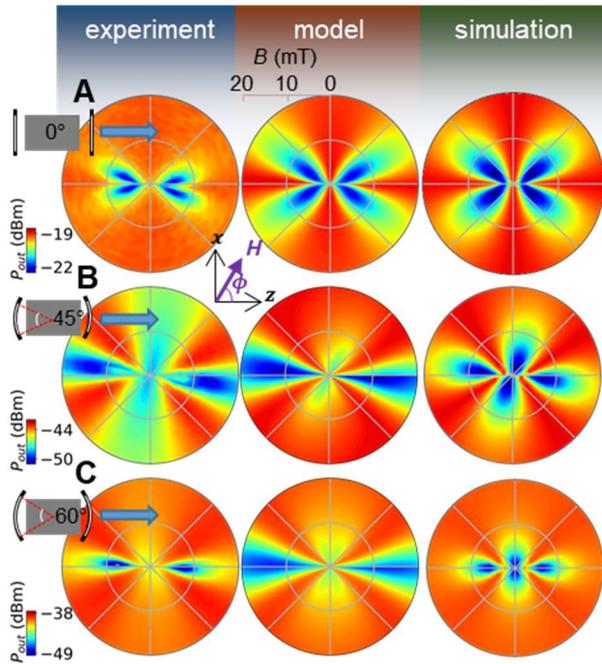

FIG. 2 Magnetic absorption of RF acoustic power in the linear regime for **A** $P_{in} = 2$ dBm, $\theta = 0°$, **B** $P_{in} = 2$ dBm, $\theta = 45°$, and **C** $P_{in} = 0$ dBm, $\theta = 60°$ devices. From left to right: experiment, model, and simulation.

However, the most important result is the increase in contrast in the low power linear regime from 2.8 dB/mm with straight IDTs $\theta = 0°$, to 9.3 dB/mm for the $\theta = 60°$ arc IDT, a value that is unprecedented for Ni at sub-GHz frequencies. The broadening of the acoustic excitation from a single SAW k-vector direction to a wide range from -30 to 30° increases interaction with the magnetic system. This increase in contrast stemming from a simple IDT design change has important implications for device design and multiferroic transduction studies, as it represents a dramatic improvement in acoustic-magnetic energy conversion efficiency. Similarly, in the same geometry with a smaller magnet (Supplemental Material), contrast was improved from 3.0 to 6.9 dB/mm by changing from the straight to curved IDT design. There, the overall effect is lower because of a smaller interaction length of the Ni thin film, but the qualitative improvement is nearly the same.

The nonlinear power dependence behavior of this device geometry is shown in FIG. 3. Our earlier publication has experimentally shown the nonlinear power dependence in traditional ADSWR devices [28]. However, this effect was extremely weak compared to our current observations in this study.

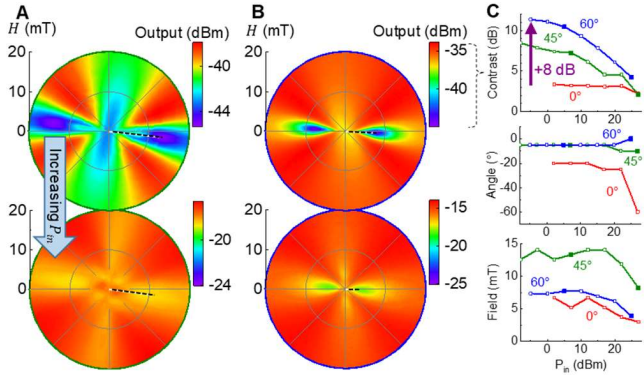

FIG. 3 Magnetic absorption of RF acoustic power for **A** 45°, and **B** 60° devices. Input power is increased from the linear (top) to nonlinear (bottom) regimes. Scale bars have the same range on top and bottom to highlight the differences. **C** Input power dependence of the contrast, and the angle and field at which maximum absorption occurs. The highlighted data points indicate the results from **A** and **B**.

In FIG. 3, the salient features of the RF input power nonlinearity on ADSWR absorption behavior in these devices are highlighted. Namely, we show that three measurable metrics (contrast, and resonant angle and field, shown in FIG. 3C) demonstrate only weak input power-dependence in the straight-IDT device over the power range explored. However, with 45 and 60° IDT arcs, clear nonlinearity is achievable even in the 0 dBm (mW) power regime, which manifests in a strongly sublinear contrast over the entire range of input powers. We conclude that the focusing of the IDTs is necessary to achieve nonlinear power dependence in this experimental range: Both 45 and 60° devices show this similarly, but that for 60° is more pronounced. This nonlinear behavior is preserved even with a smaller magnet (Supplemental Material), although the overall absorption contrast is weaker. This finding has profound implications for future studies, as it shows that novel physical behavior can be observed with modest experimental equipment, facilitating unique nonlinear device concepts. Additionally, there is potential for further improvement, for example using improved impedance matching at higher frequencies resulting in lower insertion loss, and compensation of slowness anisotropy in the piezoelectric material [31]. When we track the minimum for the lobe in the lower right quadrant of the polar ADSWR plots, we see in FIG. 3C that it begins to shift in angle for all three devices only at the highest powers. The shifts are small relative to the experimental resolution (sweeps were recorded in increments of 5°). More pronounced is the shift to lower field magnitude in all three devices. Although contrast is nonlinear even as low as 0 dBm, the position of the minimum is fairly constant in all three cases until around 15-20 dBm.

As shown in the second column of FIG. 2, we model the expected symmetry of the magnetoacoustic interaction based on the model devised by Dreher et al. [30], taking dipolar spin-wave interaction into account [19]. The model is restricted to the linear magnetoacoustic interaction regime; its results are thus independent of the actual amplitude of the SAW but reflect the symmetry of the acoustic wave [20]. We employ the coordinate system of FIG. 1. The straight IDT shown in FIG. 1A is assumed to generate a plane SAW propagating along the $z$ direction with dominant strain component $\varepsilon_{zz}$. To model the focusing effect of the FIDT in FIG. 1B phenomenologically, we assume concentric waves with displacement $u_r = \frac{A}{r}e^{ikr}$. The strain components that enter our model are given by $\varepsilon_{xx} = \frac{\partial u_x}{\partial x}, \varepsilon_{zz} = \frac{\partial u_y}{\partial y}$ and $\varepsilon_{xz} = \frac{1}{2}\left[\frac{\partial u_x}{\partial z} + \frac{\partial u_z}{\partial x}\right]$. We now average the absolute value of these strains within the Ni film, assuming vanishing strains outside of the cone angle $\theta$ defined by the FIDT design. This results in non-vanishing strain components $\varepsilon_{xx}$ and $\varepsilon_{xz}$ for the FIDT designs. To match the simple model to the experimental data we must adjust the ratio of the three strain components, generally requiring larger $\varepsilon_{xx}$ and $\varepsilon_{xz}$ than expected. This is indicative of the fact that the real wave pattern is more complex due to the anisotropic dispersion. We however still observe that increase of the arc angle $\theta$ results in more pronounced contributions from $\varepsilon_{xx}$ and $\varepsilon_{xz}$, as expected.

The micromagnetic simulations solving the full LLG equation were performed using MuMax3 [32] and the software platform Aithericon [33]. The parameters used for the simulations can be found in the Supplemental Material. The effect of the SAW was considered by including the magneto-elastic field generated by a uniform strain oscillation at the frequency of the applied SAW. A more complete model would take into account nonuniformity of the strain waves leading to spin wave excitation. However, using our analytical model, we demonstrate in the Supplemental Material that $k$ is small enough that it has little impact on the results and therefore we set k=0 in the micromagnetic simulations to lower the computation cost.

The different IDT geometries were considered by different contributions of the normal and shear strain components entering the calculation of the magneto-elastic fields like the analytical modelling.

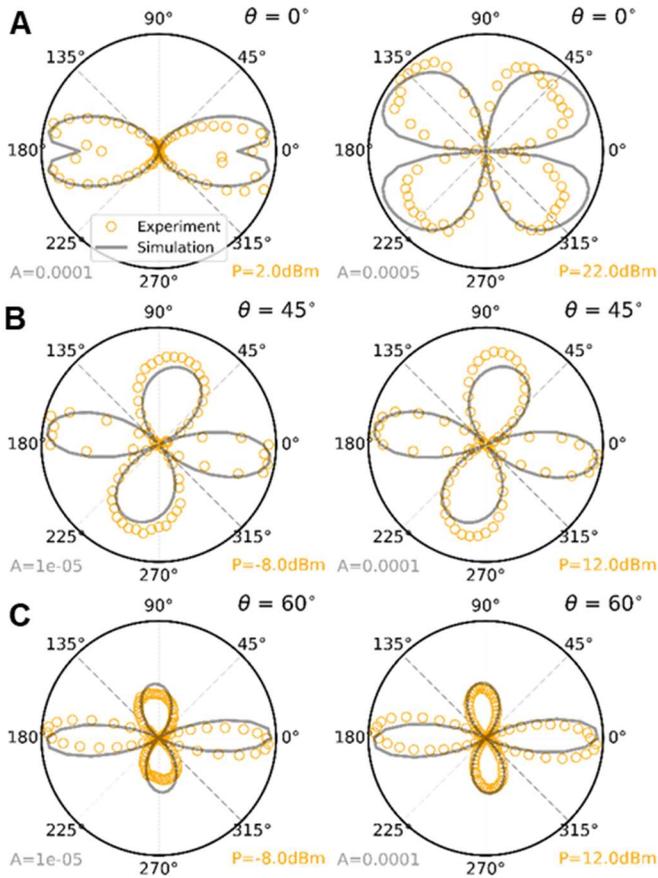

FIG. 4 Angle dependence of normalized absorbed power at 10 mT for experimental results (orange circles) at **A** $\theta = 0°$, **B** $\theta = 45°$, and **C** $\theta = 60°$, and simulations (gray curves). Left: Low power (linear regime). Right: High power (nonlinear regime).

The nonlinear power dependence behavior of the ADSWR device was also investigated by micromagnetic simulations. In FIG. 4 we show the angle dependence of the absorbed power in the magnetic system at a fixed magnetic bias field strength of 10 mT for low (linear) and high (nonlinear) acoustic power. We plot the normalized absorbed power from the experiment and the normalized excited spin-wave intensity from the simulations. In the steady state, the excited spin-wave intensity is proportional to the acoustic power absorbed by the magnetic system. Even though our simulations neglect the finite wavelength of the SAW and model the different IDT shapes based only on a change in the distribution of the normal and shear strain components, they show a good agreement with the experimental results. In FIG. 4A, for $\theta = 0°$ we see linear excitation with strain amplitude of $A = 100 \times 10^{-6}$, and the behavior only becomes nonlinear when the strain amplitude reaches $A = 500 \times 10^{-6}$. The corresponding experimental powers are 2 dBm (linear) and 22 dBm (nonlinear). In FIG. 4B and C ($\theta = 45°$ and $60°$), excitation is linear for strain amplitude of $A = 10 \times 10^{-6}$. In contrast with the $\theta = 0°$ case, here the strain amplitude only needs to reach $A = 100 \times 10^{-6}$ for the result to become nonlinear. The corresponding experimental power for Fig. 4B and C is -8 dBm (linear) to 12 dBm (nonlinear).

## Conclusions and Outlook

In this study, we have expanded on the active topic of acoustically driven spin waves by exploring the underutilized parameter of the input power. One major finding of our work is that broadening the range of excited SAW k-vector angles to tens of degrees dramatically improves their coupling to spin waves, with as much as 6.5 dB increase in efficiency even at the relatively low frequency of 860 MHz, which is far from optimal for spin waves. Moreover, we find that the symmetry of the magnetoacoustic interaction can be tuned by IDT design due to the concomitant control over SAW strain components. Sharper resonances with greater overall efficiency will lead to optimized ADSWR-based devices and further experimental results. For our second major advancement we have shown that focused interdigitated transducers have the potential to access a new area of study in nonlinearity that was previously difficult to reach. Required RF input powers are reduced from watts to milliwatts and may see further improvement with optimization of insertion loss and slowness anisotropy compensation. This opens an entire spectrum of novel nonlinear magnetoacoustic behavior for future study.

## Acknowledgements

This work is partially supported by the Air Force Office of Scientific Research (AFOSR) Award No. FA955023RXCOR001, and by the Deutsche Forschungsgemeinschaft (DFG, German Research Foundation) - TRR 173 - 268565370" (project B01).

## References

[1] S. Bandyopadhyay, J. Atulasimha, and A. Barman, "Magnetic straintronics: Manipulating the magnetization of magnetostrictive nanomagnets with strain for energy-efficient applications," *Applied Physics Reviews*, vol. 8, no. 4, p. 041323, Dec. 2021, doi: 10.1063/5.0062993.

[2] W.-G. Yang and H. Schmidt, "Acoustic control of magnetism toward energy-efficient applications," *Applied Physics Reviews*, vol. 8, no. 2, p. 021304, Jun. 2021, doi: 10.1063/5.0042138.

[3]  C. Kittel, "Excitation of Spin Waves in a Ferromagnet by a Uniform rf Field," *Phys. Rev.*, vol. 110, no. 6, pp. 1295–1297, Jun. 1958, doi: 10.1103/PhysRev.110.1295.

[4]  M. Weiler et al., "Elastically Driven Ferromagnetic Resonance in Nickel Thin Films," *Physical Review Letters*, vol. 106, no. 11, Mar. 2011, doi: 10.1103/PhysRevLett.106.117601.

[5]  L. Dreher et al., "Angle-dependent spin-wave resonance spectroscopy of (Ga,Mn)As films," *Phys. Rev. B*, vol. 87, no. 22, p. 224422, Jun. 2013, doi: 10.1103/PhysRevB.87.224422.

[6]  D. A. Bas, P. J. Shah, M. E. McConney, and M. R. Page, "Optimization of acoustically-driven ferromagnetic resonance devices," *Journal of Applied Physics*, vol. 126, no. 11, p. 114501, Sep. 2019, doi: 10.1063/1.5111846.

[7]  D. A. Bozhko, V. I. Vasyuchka, A. V. Chumak, and A. A. Serga, "Magnon-phonon interactions in magnon spintronics (Review article)," *Low Temperature Physics*, vol. 46, no. 4, pp. 383–399, Apr. 2020, doi: 10.1063/10.0000872.

[8]  K. Yamamoto, M. Xu, J. Puebla, Y. Otani, and S. Maekawa, "Interaction between surface acoustic waves and spin waves in a ferromagnetic thin film," *Journal of Magnetism and Magnetic Materials*, vol. 545, p. 168672, Mar. 2022, doi: 10.1016/j.jmmm.2021.168672.

[9]  M. Küß et al., "Symmetry of the Magnetoelastic Interaction of Rayleigh and Shear Horizontal Magnetoacoustic Waves in Nickel Thin Films on LiTaO3," *Phys. Rev. Applied*, vol. 15, no. 3, p. 034046, Mar. 2021, doi: 10.1103/PhysRevApplied.15.034046.

[10] M. Xu, J. Puebla, F. Auvray, B. Rana, K. Kondou, and Y. Otani, "Inverse Edelstein effect induced by magnon-phonon coupling," *Phys. Rev. B*, vol. 97, no. 18, p. 180301, May 2018, doi: 10.1103/PhysRevB.97.180301.

[11] S. Tateno, Y. Kurimune, M. Matsuo, K. Yamanoi, and Y. Nozaki, "Einstein--de Haas phase shifts in surface acoustic waves," *Phys. Rev. B*, vol. 104, no. 2, p. L020404, Jul. 2021, doi: 10.1103/PhysRevB.104.L020404.

[12] M. Rovirola Metcalfe, "Magnetoelastic effect with Surface Acoustic Waves in Nickel," Jul. 2021, Accessed: Oct. 17, 2022. [Online]. Available: http://diposit.ub.edu/dspace/handle/2445/180906

[13] X. Li, D. Labanowski, S. Salahuddin, and C. S. Lynch, "Spin wave generation by surface acoustic waves," *Journal of Applied Physics*, vol. 122, no. 4, p. 043904, Jul. 2017, doi: 10.1063/1.4996102.

[14] D. A. Bas et al., "Acoustically Driven Ferromagnetic Resonance in Diverse Ferromagnetic Thin Films," *IEEE Transactions on Magnetics*, vol. 57, no. 2, pp. 1–5, Feb. 2021, doi: 10.1109/TMAG.2020.3019214.

[15] P. J. Shah, D. A. Bas, I. Lisenkov, A. Matyushov, N. X. Sun, and M. R. Page, "Giant nonreciprocity of surface acoustic waves enabled by the magnetoelastic interaction," *Science Advances*, vol. 6, no. 49, p. eabc5648, Dec. 2020, doi: 10.1126/sciadv.abc5648.

[16] R. Verba, E. N. Bankowski, T. J. Meitzler, V. Tiberkevich, and A. Slavin, "Phase Nonreciprocity of Microwave-Frequency Surface Acoustic Waves in Hybrid Heterostructures with Magnetoelastic Coupling," *Advanced Electronic Materials*, vol. 7, no. 8, p. 2100263, 2021, doi: 10.1002/aelm.202100263.

[17] D. A. Bas et al., "Nonreciprocity of Phase Accumulation and Propagation Losses of Surface Acoustic Waves in Hybrid Magnetoelastic Heterostructures," *Phys. Rev. Applied*, vol. 18, no. 4, p. 044003, Oct. 2022, doi: 10.1103/PhysRevApplied.18.044003.

[18] M. Geilen et al., "Fully resonant magneto-elastic spin-wave excitation by surface acoustic waves under conservation of energy and linear momentum," *Appl. Phys. Lett.*, vol. 120, no. 24, p. 242404, Jun. 2022, doi: 10.1063/5.0088924.

[19] M. Küß et al., "Nonreciprocal Dzyaloshinskii–Moriya Magnetoacoustic Waves," *Phys. Rev. Lett.*, vol. 125, no. 21, p. 217203, Nov. 2020, doi: 10.1103/PhysRevLett.125.217203.

[20] M. Küß et al., "Nonreciprocal Magnetoacoustic Waves in Dipolar-Coupled Ferromagnetic Bilayers," *Phys. Rev. Applied*, vol. 15, no. 3, p. 034060, Mar. 2021, doi: 10.1103/PhysRevApplied.15.034060.

[21] X. Ding et al., "Surface acoustic wave microfluidics," *Lab Chip*, vol. 13, no. 18, p. 3626, 2013, doi: 10.1039/c3lc50361e.

[22] T. Franke, S. Braunmüller, L. Schmid, A. Wixforth, and D. A. Weitz, "Surface acoustic wave actuated cell sorting (SAWACS)," *Lab Chip*, vol. 10, no. 6, p. 789, 2010, doi: 10.1039/b915522h.

[23] D. J. Collins, Z. Ma, J. Han, and Y. Ai, "Continuous micro-vortex-based nanoparticle manipulation via focused surface acoustic waves," *Lab Chip*, vol. 17, no. 1, pp. 91–103, 2017, doi: 10.1039/C6LC01142J.

[24] L. Ren et al., "Standing Surface Acoustic Wave (SSAW)-Based Fluorescence-Activated Cell Sorter," *Small*, vol. 14, no. 40, p. 1801996, Oct. 2018, doi: 10.1002/smll.201801996.

[25] G. Destgeer, B. H. Ha, J. H. Jung, and H. J. Sung, "Submicron separation of microspheres via travelling surface acoustic waves," *Lab Chip*, vol. 14, no. 24, pp. 4665–4672, Sep. 2014, doi: 10.1039/C4LC00868E.

[26] D. J. Collins, A. Neild, and Y. Ai, "Highly focused high-frequency travelling surface acoustic waves (SAW) for rapid single-particle sorting," *Lab Chip*, vol. 16, no. 3, pp. 471–479, 2016, doi: 10.1039/C5LC01335F.

[27] W. Li, B. Buford, A. Jander, and P. Dhagat, "Writing magnetic patterns with surface acoustic waves," *Journal of Applied Physics*, vol. 115, no. 17, p. 17E307, May 2014, doi: 10.1063/1.4863170.

[28] D. A. Bas, P. J. Shah, M. E. McConney, and M. R. Page, "Optimization of acoustically-driven ferromagnetic resonance devices," *Journal of Applied Physics*, vol. 126, no. 11, p. 114501, Sep. 2019, doi: 10.1063/1.5111846.

[29] D. Labanowski, A. Jung, and S. Salahuddin, "Power absorption in acoustically driven ferromagnetic resonance," *Appl. Phys. Lett.*, vol. 108, no. 2, p. 022905, Jan. 2016, doi: 10.1063/1.4939914.

[30] L. Dreher et al., "Surface acoustic wave driven ferromagnetic resonance in nickel thin films: Theory and experiment," *Phys. Rev. B*, vol. 86, no. 13, p. 134415, Oct. 2012, doi: 10.1103/PhysRevB.86.134415.

[31] R. O'Rorke, A. Winkler, D. Collins, and Y. Ai, "Slowness curve surface acoustic wave transducers for optimized acoustic streaming," *RSC Adv.*, vol. 10, no. 20, pp. 11582–11589, 2020, doi: 10.1039/C9RA10452F.

[32] A. Vansteenkiste, J. Leliaert, M. Dvornik, M. Helsen, F. Garcia-Sanchez, and B. Van Waeyenberge, "The design and verification of MuMax3," *AIP Advances*, vol. 4, no. 10, p. 107133, Oct. 2014, doi: 10.1063/1.4899186.

[33] "Aithericon." https://aithericon.com/ (accessed Sep. 13, 2022).

# Supplemental Material

## Experiment

For polar ADSWR measurements like those in Figs. 2 and S1, a signal generator (Keysight N5171B) delivers pulsed RF to the ADSWR device, and the output is measured using a spectrum analyzer (Keysight N9000A). For the power dependence study, we discretely vary the input power from +5 to +27 dBm. Time-gating is used to isolate the signal transmitted via SAWs, which is delayed by about 1 μs compared to the EM radiative signal because of the slower velocity of SAWs. A vector electromagnet is used to sweep the angle $\phi$ and magnitude $H$ of the magnetizing field. The magnitude is swept from high (5 mT) to low (0 mT) to ensure consistency in hysteretic behavior. A vector network analyzer is used to calibrate the transmission values.

IDT patterning for metal-liftoff was completed using negative tone lift-off photoresist NR9-1000Py and a Karl Suss MA6 mask aligner contact lithography system. The Al electrode thickness is 70 nm deposited using e-beam evaporation. In the spacing between the IDTs, 20-nm-thick Ni film was patterned and deposited via standard lithography and e-beam evaporation.

## Smaller magnet

Fig. S1 summarizes ADSWR results for a smaller magnet. Here, the focal length is $F_1 = 400$ μm, and the magnet dimensions are $L_x = 400$ μm, $L_z = 612$ μm. Fig. S2 can be compared to Fig. 3C. As expected, the devices trend similarly, with little change in the control ($\theta = 0°$) device (red), and increasing power dependence for $\theta = 45°$ (green) and $\theta = 45°$ (blue).

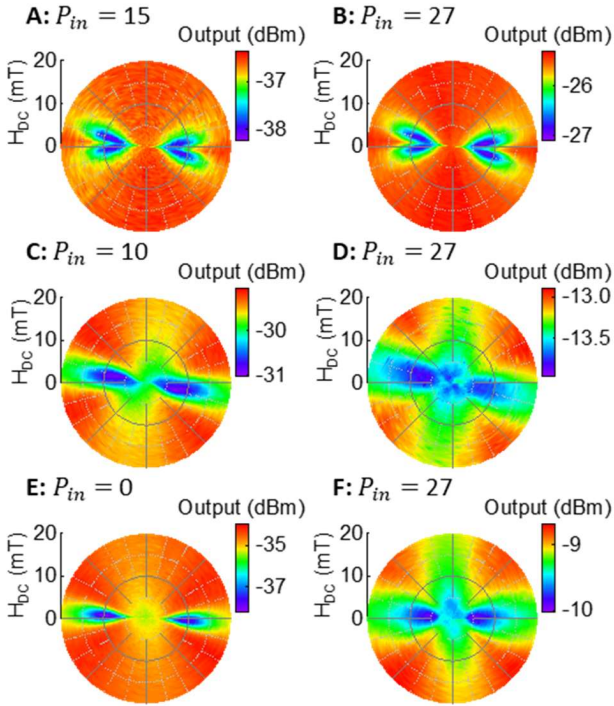

FIG. S1 ADSWR results for the smaller magnet with **A** $P_{in} = 15$ dBm, $\theta = 0°$, **B** $P_{in} = 27$ dBm, $\theta = 0°$, **C** $P_{in} = 10$ dBm, $\theta = 45°$, **D** $P_{in} = 27$ dBm, $\theta = 45°$, **E** $P_{in} = 0$ dBm, $\theta = 60°$, **F** $P_{in} = 27$ dBm, $\theta = 60°$.

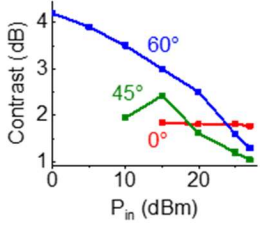

FIG. S2 Power dependent contrast for smaller magnet.

## Acoustic linearity

To confirm that the nonlinearity stems from the magnetism rather than from the acoustic waves themselves, we measured the transmission while applying a high magnetic field well above the resonance. In Fig. S3 this measurement is shown for all the devices used in this work ("S" indicates the smaller magnet and "L" the larger, and the angle refers to the arc angle $\theta$). In the double-logarithmic plot we find that all linear fits (lines) have slope≈1. This demonstrates that in the absence of magnetic interaction, the behavior is purely linear in all cases.

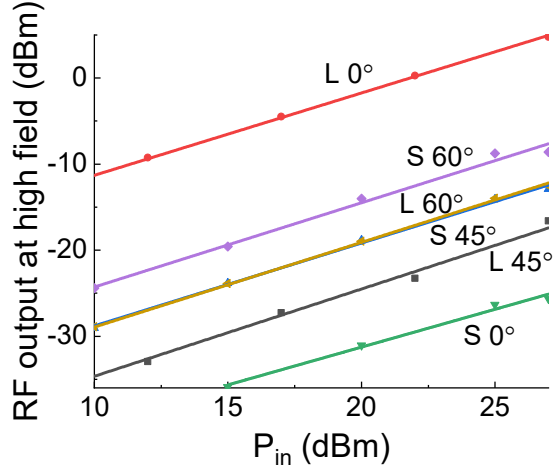

FIG. S3 SAW transmission at high applied field $H_{DC}$.

## Micromagnetic simulation parameters

The details for the micromagnetic simulations are presented below. The simulated system is a pad with the dimensions of 10.24 μm × 10.24 μm × 20 nm which is divided into 512 × 512 × 1 cells. We have used periodic boundary conditions in both x and y directions to mimic a plane film. The following parameters have been used: saturation magnetization $M_S = 470$ kA/m, exchange stiffness $A_{ex} = 11$ pJ/m, Gilbert damping parameter $\alpha = 0.3$ and the magnetoelastic coupling constants $B_1 = B_2 = 10$ MJ/m$^3$. Uniaxial anisotropy along the x-axis has been assumed with $K_{u1} = 705$ J/m$^3$. The angle between the x-axis and the magnetic field is $\phi$. The SAWs are represented by a plane wave with frequency $f$ and amplitude $A$ for the strain components $\varepsilon_{xx}$, $\varepsilon_{yy}$, and $\varepsilon_{xy}$.

$$\varepsilon_{xx} = \beta_{xx} A \sin(2\pi f t)$$
$$\varepsilon_{xy} = \beta_{xy} A \sin(2\pi f t)$$

$$\varepsilon_{yy} = \beta_{yy} A \sin(2\pi f t)$$
$$\varepsilon_{yz} = 0$$
$$\varepsilon_{xz} = 0$$
$$\varepsilon_{zz} = 0$$

for $\theta = 0, 45, 60$, $\beta_{xx} = 1, 1, 1$, $\beta_{xy} = 0, 0.7, 1$, $\beta_{yy} = 0, 0.7, 1$.

For a strain amplitude of $A = 10 \times 10^{-6}$, a linear excitation can be seen at $f$. Nonlinearity is observable with a strain amplitude of $A = 100 \times 10^{-6}$ (for $\theta = 45, 60°$) and $A = 500 \times 10^{-6}$ (for $\theta = 0°$).

The dynamic components of the magnetization are collected over a period of 300 ns and recorded in 10 ps intervals. The data are analyzed using a fast Fourier transformation in space and time and the intensity is extracted for frequency $f$.

**Analytical model**

We use the analytical model for the SAW-spin-wave interaction described in Ref. [19].

The model is based on solving the Landau-Lifshitz-Gilbert equation
$$\frac{\partial \boldsymbol{m}}{\partial t} = -\gamma \mu_0 (\boldsymbol{H}_{\text{eff}} + \boldsymbol{h}) \times \boldsymbol{m} + \alpha\, \boldsymbol{m} \times \frac{\partial \boldsymbol{m}}{\partial t}$$
with an effective field $\boldsymbol{H}_{\text{eff}}$ describing the magnetic system without considering the SAW and a driving field $\boldsymbol{h}$ describing the impact of the SAW on spin dynamics.

For solving the LLG equation, we consider a cartesian 123-coordinate system with the 3-axis aligned along the in-plane equilibrium orientation of the magnetization. The 1-axis points out-of-plane, and the 2-axis is in-plane orthogonal to the 3-axis. This 123-coordinate system and the necessary transformations between the laboratory xyz-system discussed in the main text are described in detail in Ref. [30]. The effective field in the 123-coordinate system is given by
$$\boldsymbol{H}_{\text{eff}} = \boldsymbol{H} + H_u (\boldsymbol{m} \cdot \boldsymbol{u}) - M_s \begin{pmatrix} G_0 m_1 \\ (1 - G_0) m_2 \sin^2(\phi_0) \\ 0 \end{pmatrix} - \frac{2 A_{\text{ex}}}{\mu_0 M_s} k^2 \begin{pmatrix} m_1 \\ m_2 \\ 0 \end{pmatrix}$$

where the terms are, in order, the external magnetic field, a uniaxial anisotropy field $H_u$ along the unit vector $\boldsymbol{u} \parallel \boldsymbol{x}$, the dipolar spin-wave interactions with $G_0 = (1 - e^{-|k|d})/(|k|d)$ and the exchange spin-wave interactions. $M_s = 470$ kA/m is the saturation magnetization, $A_{\text{ex}} = 11$ pJ/m is the exchange constant. $\phi_0$ is the angle between the equilibrium $\boldsymbol{m}$ orientation and the z-axis (see main text), $m_1$ and $m_2$ are the dynamic magnetization components with $m_1, m_2 \ll 1$ and $m_3 = 1$.

The two components of the SAW driving field $\boldsymbol{h}$ that are orthogonal to the equilibrium $\boldsymbol{m}$ direction are given in the 123-coordinate system and assuming the non-zero strains $\varepsilon_{zz}, \varepsilon_{xx}$ and $\varepsilon_{xz}$ (see main text) by
$$\boldsymbol{h} = \begin{pmatrix} h_1 \\ h_2 \end{pmatrix} = \frac{b}{\mu_0} \begin{pmatrix} 0 \\ (\varepsilon_{zz} - \varepsilon_{xx}) \sin(2\phi_0) - 2\, \varepsilon_{xz} \cos(2\phi_0) \end{pmatrix}$$

For the straight IDT, we set $\varepsilon_{zz} = 1$ and $\varepsilon_{xx} = \varepsilon_{xz} = 0$ to model a plane wave. For the 45° IDT we use $\varepsilon_{zz} = 1$ and $\varepsilon_{xx} = \varepsilon_{xz} = 0.7$ and for the 60° IDT we use $\varepsilon_{zz} = \varepsilon_{xx} = \varepsilon_{xz} = 1$ to model the concentric waves with a given IDT opening angle. For each $\mathbf{H}$, we first numerically determine the equilibrium $\mathbf{m}$ orientation $\phi_0$ by free energy minimization and then numerically solve the LLG equation with the harmonic ansatz $m_i = m_{i,0}\, e^{-i}$ . From this, we obtain $m_{i,0}$ as a function of $\mathbf{H}$. In a final step, we calculate the absorbed power $P_{abs} \propto \text{Im}\{\mathbf{h} \cdot \mathbf{m_0}\}$ as described in detail in Ref. [19].

**Uniform and non-uniform strain and spin dynamics**

In our experiments, we use SAWs with frequency of 860 MHz and wavevector $k_{SAW} \approx 1.5 \times 10^6$ m$^{-1}$. To demonstrate the impact of considering excitation of a spin-wave with $k_{SW} = k_{SAW}$ compared to the simplification of FMR excitation by uniform strain we performed analytical model calculations with $k = k_{SAW}$ and $k = 0$ with the results shown in Fig. S4. While some quantitative changes are observed, the overall salient features remain identical. We thus carried out the power-dependent micromagnetic simulations by assuming uniform strain to drastically reduce the required computation time.

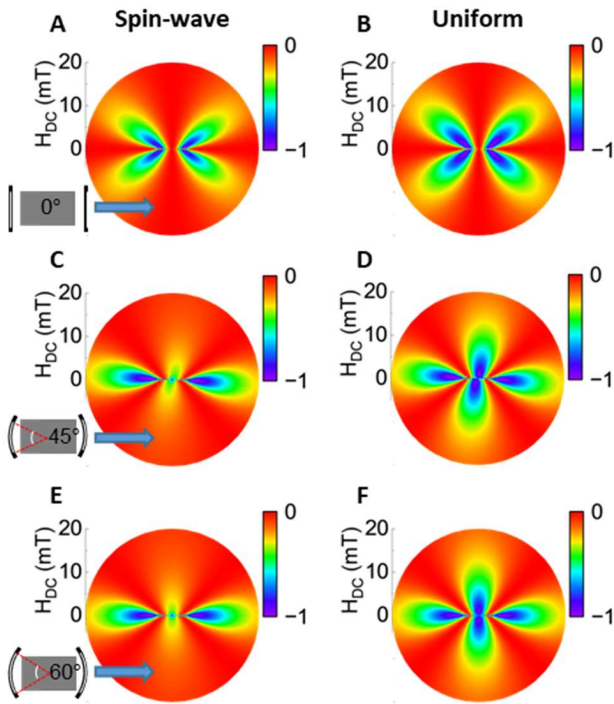

FIG. S4 Analytical model calculations considering the non-zero wavevector leading to spin-wave excitations (left column) and assuming simplification of uniform strain and spin dynamics (right column). The colorbar shows the output in arbitrary logarithmic units.